\documentstyle[aas2pp4]{article}

\def\lesssim{\,
\lower2truept\hbox{${<\atop\hbox{\raise4truept\hbox{$\sim$}}}$}\,}

\def\gtrsim{\,\lower2truept\hbox{${> \atop\hbox{\raise4truept\hbox{$\sim$}}}$}\,}

\newcommand{\mc}{{mc}}

\slugcomment{Submitted to The Astrophysical Journal, 1998}

\begin{document}

\title{Modelling the effects of dust on galactic SEDs from the UV
to the millimeter band}

\author{Laura\ Silva}
\affil{International School for Advanced Studies, SISSA, Trieste, Italy}

\author{Gian Luigi\ Granato, Alessandro Bressan}
\affil{Osservatorio Astronomico di Padova, Padova, Italy}

\and

\author{Luigi\ Danese}
\affil{International School for Advanced Studies, SISSA, Trieste, Italy}

\begin{abstract}

We present models of photometric evolution of galaxies in which the effects
of a dusty interstellar medium have been included with particular care. A
chemical evolution code follows the star formation rate, the gas fraction
and the metallicity, basic ingredients for the stellar population synthesis.
The latter is performed with a grid of integrated spectra of simple stellar
populations (SSP) of different ages and metallicities, in which the effects
of dusty envelopes around asymptotic giant branch (AGB) stars are included.
The residual fraction of gas in the galaxy is divided into two phases: the
star forming molecular clouds and the diffuse medium. The relative amount is
a model parameter. The molecular gas is sub--divided into clouds of given
mass and radius: it is supposed that each SSP is born within the cloud and
progressively escapes it. The emitted spectrum of the star forming molecular
clouds is computed with a radiative transfer code. The diffuse dust emission
(cirrus) is derived by describing the galaxy as an axially symmetric system,
in which the local dust emissivity is consistently calculated as  a function
of the local field intensity due to the stellar component. Effects of very
small grains, subject to temperature fluctuations, as well as polycyclic
aromatic hydrocarbons (PAH) are included.

The model is compared and calibrated with available data of normal and
starburst galaxies in the local universe, in particular new broad--band and
spectroscopic ISO observations. It will be a powerful tool to investigate
the star formation, the initial mass function (IMF), supernovae rate (SNR)
in nearby starbursts and normal galaxies, as well as to predict the
evolution of luminosity functions of different types of galaxies at
wavelengths covering four decades.

\keywords{
dust, extinction ---
galaxies: ISM ---
galaxies: spiral --- 
galaxies: starburst ---
infrared: galaxies ---
radiative transfer
}
\end{abstract}

\section{Introduction}

Observations performed in the last decade or so, particularly in the IR
regime, have clearly demonstrated that dust is one of the most important
components of the interstellar medium (ISM), containing a large fraction of
heavy elements ejected from stars. This paper is mainly concerned with the
influence of dust grains on the transfer of radiation emitted by stellar
systems. Basically dust absorbs and scatters photons, mostly at wavelengths
$\lesssim 1$ $\mu$m and returns to the radiation field the subtracted energy
in the form of IR photons. The resulting spectral energy distribution (SED)
is often substantially changed and in many relevant cases radically
modified.

Not surprisingly, dust reprocessing of the optical--UV photons emitted by
stars into the infrared regime turns out to be particularly severe in
galaxies undergoing massive episodes of star formation, which preferentially
occur within dense molecular clouds. Indeed dust could affect galaxy
evolution because it modifies the physical and chemical conditions of the
ISM. In particular star formation is at least favored by the presence of
dust, which shields dense clouds from stellar UV radiation and keeps them to
temperatures low enough to allow the onset of gravitational instability.

The SED of normal star--forming disk--like galaxies is also affected by
dust, mainly associated with the diffuse ISM and with the envelopes of
evolved asymptotic giant branch (AGB) stars. As for early type systems, it
has been often suggested that the first episode of massive star--formation
could be essentially hidden to optical searches due to dust reprocessing,
since their ISM might have been metal enriched on a very short timescale,
the lifetime of the first generation of massive stars (Franceschini et al.\
1994 and references therein; Cimatti et al.\ 1997). Even local quiescent
elliptical galaxies exhibit infrared emission (e.g.\ Jura 1986; Bally \&
Thronson 1989; Knapp et al.\ 1989; Roberts et al.\ 1991) and visual dust
obscuration (e.g.\ Veron--Cetty \& Veron 1988; Kim 1989; Goudfrooij et al.\
1994), arising in part from dust grains continuosly formed in dusty outflows
from AGB stars (Tsai \& Mathews 1996; Bressan, Granato, \& Silva 1998), as
well as molecular line emission (e.g.\ Gordon 1990; Roberts et al.\ 1991;
Lees et al.\ 1991). Overall, a direct comparison of the luminosity functions
of galaxies in the optical and in the far--IR shows that, locally, about
30\% of starlight is dust--reprocessed.

\nocite{FMD94,Jur86,BaT89,KGK89,RHB91,VeV88,Kim89,TsM96,Gor90,LKR91,GDH94}

At least three different dusty environments must be taken into account in
order to properly understand the UV to sub--mm properties of galaxies: (i)
dust in interstellar H I clouds heated by the general interstellar radiation
field (ISRF) of the galaxy (the `cirrus' component), (ii) dust associated
with star forming molecular clouds and H II regions and (iii) circumstellar
dust shells produced by the windy final stages of stellar evolution. These
environments have different importance in different galactic systems at
various evolutionary stages.

Despite this, in many papers dealing with the spectrophotometric evolution
of galaxies the radiative processes occurring in a dusty interstellar
medium were originally neglected (e.g.\ Bruzual \& Charlot 1993 and
Leitherer \& Heckman 1995, the former the most adopted one to interpret
data on different galaxy types, the latter appositely built for starburst
galaxies). Indeed their codes are nowadays commonly used together with
some prescription to approximate the effects of the ISM. In other cases
(e.g.\ Guiderdoni \& Rocca--Volmerange 1987; Mazzei, De Zotti, \& Xu 1994;
Lan\c{c}on \& Rocca--Volmerange 1996; Fioc \& Rocca--Volmerange 1997), the
effects of dust are included only partially and/or with substantial
simplifications. For instance often extinction is considered but thermal
reradiation is not, or scattering is neglected, or not all the relevant
dusty environments are considered, or optically thin emission is assumed,
or a unrealistic geometry is adopted. Those works facing a complete
computation of the radiative transfer through a dusty medium, do not
include it in the more general framework of galaxy evolution, but are
instead interested either in the interpretation of single objects (e.g.\
Kr\"ugel \& Siebenmorgen 1994) or in understanding the effects of dust
geometry on the extinction of the emerging spectrum (e.g.\ Bruzual,
Magris, \& Calvet 1988; Witt, Thronson, \& Capuano 1992; Cimatti et al.\
1997; Wise \& Silva 1996; Bianchi, Ferrara, \& Giovanardi 1996; Gordon,
Calzetti, \& Witt 1997).

\nocite{BrC93,LeH95,GuR87,MDX94,FiR97,LaR96}
\nocite{KrS94,BMC88,WTC92,CBF97,WiS96,BFG96,GCW97}

The inclusion of dust effects into spectrophotometric codes leads to
many difficulties. While the integrated photometric properties of an
hypothetical dust--free galaxy are geometry independent, dust
introduces a strong dependence on the distribution of both stars and
ISM. Moreover the optical properties of dust grains and  their
dependence on environmental conditions have not yet been fully
explored and understood. The ensuing uncertainty may only be
parametrized to some extent. The effects of scattering, absorption and
emission of grains, which may not be in radiative equilibrium with
the radiation field, complicate the solution of the
integro--differential transfer equation in a complex geometry.

Nonetheless a more realistic and complete computation of the
spectrophotometric galaxy models including dust is required, since
neglecting  the complexity of the dust effects can lead to erroneous
estimates of many interesting quantities such as the star formation rate
(SFR). Also, the estimate of the age of a galaxy through the fit of its
broadband SED is hampered by the degeneracy between the colors of an old
galaxy and those of an extinguished young galaxy (e.g.\ Cimatti et al.\
1997), further complicating the well known age--metallicity degeneracy.

\nocite{CBF97}

We have therefore developed chemo--spectrophotometric self--consistent
galactic models including dust. Three different dusty environments are
considered: envelopes of AGB stars, molecular star--forming clouds and
diffuse ISM; the dust model is comprehensive of normal big grains, small
thermally fluctuating grains and polycyclic aromatic hydrocarbons (PAH)
molecules; for each grain family, the appropriate computation of
absorption and reemission is performed. The models are suited to simulate
galaxies of any Hubble type, in different evolutionary stages, since we
include the possibility of different geometrical distributions both for
stars and gas. The complete radiative transfer equation is numerically
solved whenever necessary.

The many parameters and uncertainties introduced by the presence of dust can
be constrained only by means of the multiband approach pursued in this
paper, where the spectrophotometric properties of galaxies are consistently
reproduced from the UV to the sub--mm, i.e.\ taking into account extinction
at short wavelengths and the consequent thermal emission in the IR regime.
The combination of observations from HST, Keck, ISO and ground--based
optical, IR and sub--mm telescopes already provided several objects at
significant redshift with spectral information on this large $\lambda$
range. Their number will increase after the completion of ISO surveys, and
will burst when SIRFT, FIRST, Planck Surveyor and NGST will operate.

In this paper we present the model and we use it to study local galaxies. In
a forthcoming paper we will exploit it to study the cosmic evolution of
different types of galaxies.

\section{Model description}
\label{secmod}

We divide the problem of estimating the SED of a galaxy at age
$t_G$, including dust effects, in two steps: (1) the history  of the
star formation rate $\Psi(t)$, of the initial mass function (IMF), of
the metallicity $Z(t)$ and of the residual gas fraction $g(t)$ is
determined and then (2) the integrated SED of the galaxy is predicted
taking into account all the stars and the gas present at $t_G$. When
the effects of dust are neglected, step (2) simply involves a sum of
all the spectra of stars. In our case we have instead to introduce a
specific geometry for gas and stars, and then to compute the radiative
transfer of the radiation emitted by the stars and the dust.

The purpose of the present work is mainly to develop a general
procedure for step (2). Step (1) includes a number of possible choices
which can be easily interfaced to our code in order to describe the
evolution of the SED.

\subsection{Chemical evolution} 
\label{seccheevo}

The chemical evolution is a preliminary process in our code. We summarize
here the main aspects of the code, which basically follows the guidelines
given by Tantalo et al.\ (1996).

\nocite{TCB96}

The code describes one--zone open models including the infall of primordial
gas ($\dot M(t) \propto \exp(-t/t_{inf})$), in order to simulate the collapse
phase of galaxy formation and,  when required by the astrophysical situation
under study, galactic winds.

The adopted SFR is a Schmidt--type law, i.e. proportional to some power
(between 1 and 2) of the available gas mass: $\Psi(t)=\nu \,
M_{g}(t)^{k}$. For starburst galaxies we add to the general smooth SFR one
or more bursts of star formation. As for the IMF we used the usual Salpeter
law: $\Phi(M)\propto M^{-x}$, with $x=2.35$. The upper limit of the mass
range is fixed to $100$ $M_\odot$.

This kind of chemical evolution models has been successfully tested
against nearby spheroidal galaxies, and also well reproduces the
properties of different types of galaxies with appropriate choices of
the parameters (e.g. a quiet spiral--type evolution is well mimicked
with a low value of $\nu$ and/or a long infall time scale, see
Matteucci 1996 for a thorough review).

\nocite{MatF96}

\subsection{Synthesis of starlight spectrum (including dusty 
envelopes of AGB stars)}
\label{secsyn}

The library of isochrones for the simple stellar populations (SSP), the
building--blocks of galaxy models, is based on the Padua stellar models
(Bertelli et al.\ 1994),
with a major difference, consisting in the computation of the effects of
dusty envelopes around AGB stars. The SSPs span a wide range in age, from 1
Myr to 20 Gyr, and in metallicity, $Z=0.004,0.008,0.02,0.05,0.1$, to
realistically reproduce the mix of age and composition of the stellar
content of galaxies. The inclusion of a wide range of metallicity  is
imposed by the aim of comparing model estimates to the excellent data
available even for young high redshift galaxies.

The spectral synthesis technique, for the starlight alone, consists
in summing up the spectra of each stellar generation, provided by
the SSP of the appropriate age and metallicity, weighted by the SFR
at time of the stars birth (e.g.\ Bressan, Chiosi, \& Fagotto 1994).

Effects of dust in the envelopes of evolved stars are usually neglected in
the synthesis of a composite population. This is not justified for
intermediate age population clusters, whose brightest tracers are AGB stars,
and/or in a wide--wavelength synthesis approach. To overcome this
limitation, we computed new isochrones and SSPs in which, along the AGB, a
suitable dusty envelope is assumed to surround the star. For this envelope
the radiative transfer is solved by means of the code described by Granato
\& Danese (1994). The envelope parameters, in particular its optical
thickness, which is straightforwardly linked to the mass-loss rate and to
the expansion velocity, are derived as a function of basic stellar
parameters (mass, luminosity, radius, and metallicity) combining
hydrodynamic model results and empirical relations. A detailed description
of the adopted procedure can be found in Bressan et al.\ (1998).

\nocite{GrD94,BGS98}

\subsection{Geometry}
\label{secgeo}

\begin{figure}
\epsscale{1.0}
\plotone{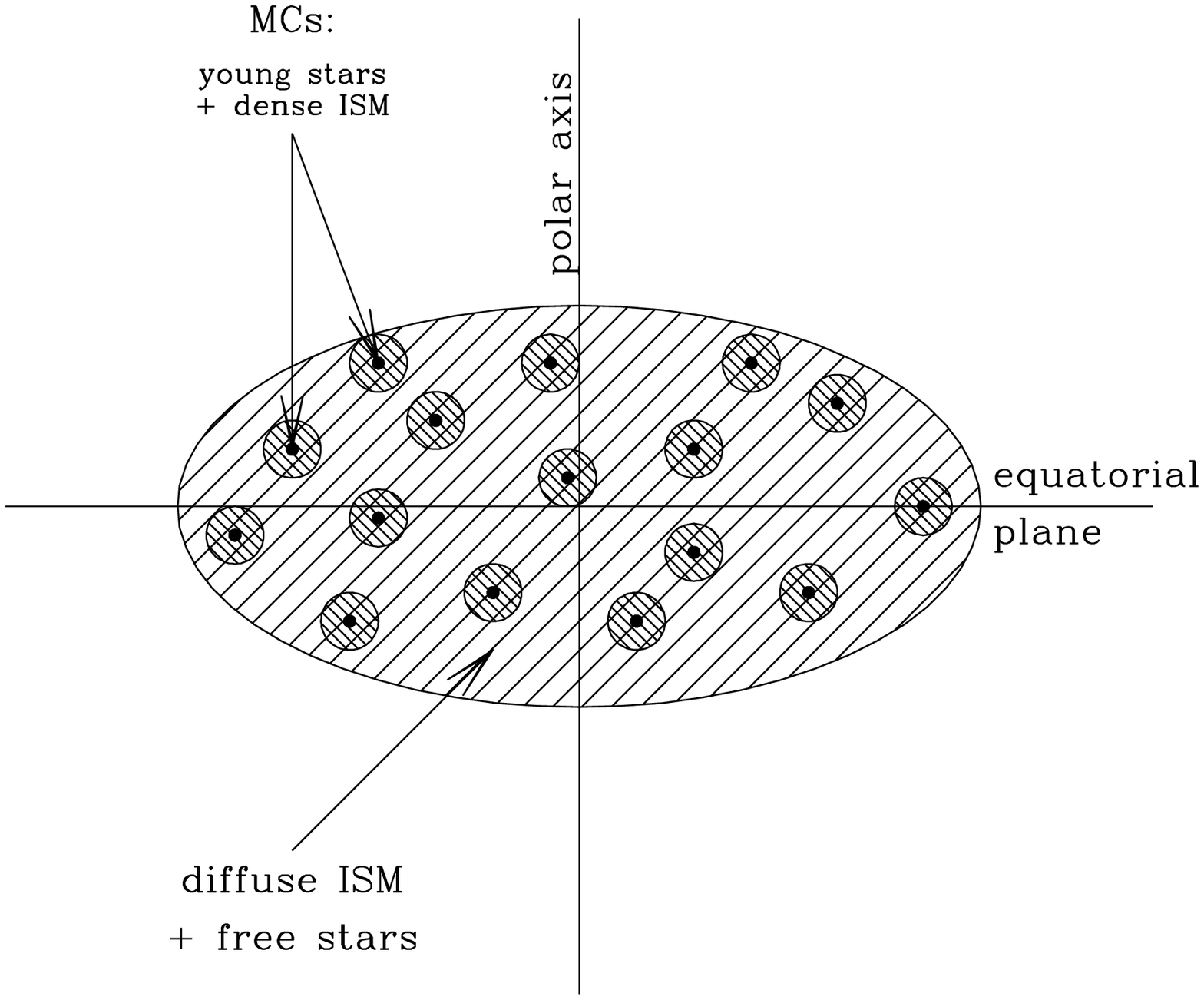}
\caption{Scheme of the components  
included in our model computations and their adopted geometry.}
\label{figcar}
\end{figure}

As already pointed out, the dust effects on SEDs depend on the relative
distribution of stars and dust. The assumed geometry is sketched in Fig.
\ref{figcar}. The galaxy is approximated as a system having azimuthal
symmetry as well as planar symmetry with respect to the equatorial plane. We
take into account three components: i) star forming molecular clouds
complexes, hereafter MCs, comprising dusty gas in a dense phase, H II
regions and very young stars embedded in it (young stellar objects YSO); ii)
stars already escaped from these dense clouds (henceforth {\it free stars});
iii) diffuse gas (cirrus).

We work in spherical coordinates $(r,\theta,\phi)$. The densities of the
three components $\rho_\mc$, $\rho_*$, and $\rho_c$ respectively, depend on
$r$ and $\theta$ through analytical laws. In the following, unless
otherwise explicitly stated, $\rho_\mc$ and $\rho_*$ have identical spatial
dependence. In order to describe disk--like galaxies, we use a double
exponential of the distance from the polar axis $R = r \sin \theta$ and
from the equatorial plane $z = r \cos \theta$:
\begin{equation} 
\rho =
\rho_o \, \exp(- R/R_d) \,  \exp(- |z|/z_d) \: . \label{equdis}
\end{equation} 
In the code the scale lengths $R_d$ and $z_d$ can be independently set for the
three components, but in the models presented in this paper (Sec.\
\ref{seccompa}) we simply adopt identical values for them. In other words we use
two adjustable parameters when describing disk--like galaxies:
$R_d^*=R_d^\mc=R_d^c$ and $z_d^*=z_d^\mc=z_d^c$. Observations within our Galaxy
suggest $z_d^{\star}\sim 0.35$ kpc. As a reference the {\it e}-folding scale
length is related to the absolute magnitude by
\begin{equation} 
\log_{10}(R_d/{\rm kpc}) \sim -0.2 M_B-3.45 \: , 
\label{equscale}
\end{equation} 
(Im et al.\ 1995).

In the case of spheroidal systems we adopt for both stars and dust
spherical symmetric distributions with King profile:
\begin{equation} \rho =
\rho_o (1+(r/r_c)^2)^{-\gamma} \: ,
\label{equbul} 
\end{equation} 
extended up to the tidal radius $r_t$. This truncation radius is required in
King models since $M(r)$, the mass contained within $r$, diverges as $r
\rightarrow \infty$. Our results are not very sensitive to its precise
choice, thus we simply adopted the standard value $\log (r_t/r_c)=2.2$. As
for the stellar component we simply set $\gamma = 3/2$. It has been
suggested that the core radius $r_c$ correlates with the luminosity
(Bingelli, Sandage, \& Tarenghi 1984):
\begin{equation}
\log_{10}\left( \frac{r_c}{\rm{kpc}} \right)\sim 
\left\{ \begin{array}{ll}
-0.3\, (M_B+22.45) & \textrm{if $M_B\le-20$}\: ,\\
-0.1\, (M_B+27.34) & \textrm{if  $M_B>-20$} \: .
\end{array}
\right. 
\end{equation}

The distribution of dust in spheroidal systems is poorly known, but it has been
suggested that $\rho_{\rm dust} \propto \rho_{\rm stars}^n$ with $n \simeq
1/2\div 1/3$ (Froehlich 1982; Witt et al.\ 1992; Wise \& Silva 1996), i.e.\
$\gamma \simeq 0.5\div 0.75$. In other words the dust distribution seems to be
less concentrated than that of stars. In particular an additional cool
component located in the outer parts has been invoked to explain IRAS
observations (Tsai \& Mathews 1996).

The volume emissivity (ergs cm$^{-3}$ s$^{-1}$ ster$^{-1}$ \AA$^{-1}$)
of the galaxy  at each point is the sum of three terms, arising
respectively from the three components listed above:
\begin{equation}
j_\lambda= j_\lambda^\mc + j_\lambda^* + j_\lambda^c \: .
\end{equation}
In the following sections we will describe how these quantities, as
well as the specific flux measured by an external observer, are
computed. A preliminary stride is to define the optical properties
of dust.

\subsection{The dust model}
\label{secdus}

Once the geometrical arrangement is specified, the effects of dust
on radiative transfer depend on physical and chemical properties of
grains, which affect the way they absorb and emit photons. It is expected
and observationally well established that these properties are functions of
the particular environment in which grains happen to live. Several
populations may be distinguished (see Dorschner \& Henning 1995 for a review
): (i) stellar outflow dust (further subdivided into the two chemical
subgroups of carbon rich and oxygen rich dust); (ii) dust in the diffuse
ISM; (iii) dust in molecular clouds and dust around YSOs.  As for stellar
outflow dust, we refer to the treatment of Bressan et al.\ (1998) mentioned
above. Here we discuss our choices for the other two dusty components of the
model galaxy.

\nocite{Fro82,DoH95,BGS98}

Although a lot of efforts have been devoted to derive a so called {\it
standard} model for dust in the diffuse ISM, its precise composition remains
controversial. Significant clues can be derived from observations. The
prominent features observed in its spectrum at 9.7 and 18 $\mu$m indicate
the presence of silicate grains. By converse, the origin of the 2175 \AA\
feature is somewhat debated, though graphite still stands as the most
attractive solution (see Mathis 1990). The emission bands in the 3-13
$\mu$m region indicate the presence of PAHs (Puget, L\'eger, \& Boulanger
1985), which together with small thermally fluctuating grains contribute
also to produce the warm mid--IR cirrus emission. Therefore we included
these components in our diffuse dust model. We adopted the same mixture even
for MCs, for which the information are scanty, only decreasing the fraction
of PAH molecules (see below).
 
\nocite{Mat90} 
 
The optical properties of silicate and graphite grains have been
computed by Laor \& Draine (1993) for spherical shapes using Mie
theory, the Rayleigh--Gans approximation and geometric optics. In
particular we used the cross sections of graphite and silicate grains
computed by B.T.\ Draine for 81 grain sizes from 0.001 to 10 $\mu$m in
logarithmic steps $\delta \log a = 0.05$, and made available via
anonymous ftp at {\it astro.princeton.edu}.

\nocite{LaD93}

In order to get the overall radiative properties of the dust mixture, the
abundance and the size distribution of grains must be specified. We started from
the distribution proposed by Draine \& Lee (1984, henceforth DL), which is tuned
on the optical--UV extinction law of the galactic diffuse ISM. On the other hand
DL pointed out that their model displays significant discrepancies with
extinction data above 2 $\mu$m, and that these discrepancies are not surprising,
since observations at longer $\lambda$ tend to sample lines of sight through
denser clouds, where grain population may differ from that present in more
diffuse regions.

\nocite{DrL84}

\begin{figure}
\epsscale{1.0}
\plotone{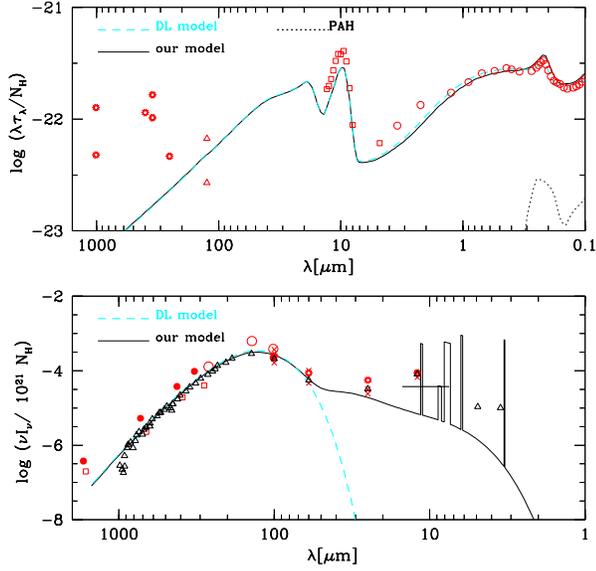}
\caption{Upper panel: the extinction curves of the dust model we
adopt for diffuse and molecular gas, and that by Draine \& Lee (1984)  
are compared with
available data. Lower panel: the
predicted emissivity of grains (ergs s$^{-1}$ cm$^{-2}$ st$^{-1}$
N$_H^{-1}$) in the local interstellar radiation field (Mathis,
Mezger, \& Panagia 1983) compared with observations toward the
galactic pole. The orizontal line around 12$\mu$m marks the flux level
obtained by convolving our expected SED with the 12$\mu$m IRAS passband.
Triangles in the lower panel are data from Dwek et
al.\ (1997), for references to other observations see Rowan--Robinson (1992,
1986)}
\label{figesci}
\end{figure}

A more severe limitation of DL model is that the mid--IR (MIR) cirrus
emission at wavelengths $\lambda \leq $60 $\mu$m is not reproduced (Fig.
\ref{figesci}), because grains are always large enough to maintain a low
temperature thermal equilibrium in the general ISRF. This emission requires
reprocessing of the field by particles, very small grains and/or PAH
molecules, reaching temperatures higher than those attained with equilibrium
heating (e.g.\ Puget et al.\ 1985; Draine \& Anderson 1985; Dwek et al.\
1997). This requirement can be matched by decreasing the lower limit of the
graphite grain distribution. Since these grains must be more numerous than
the simple extrapolation of the DL power law, a break to a steeper power law
is introduced below $a_b$. These adjustments to the DL model enhance the MIR
cirrus emission, but tend to degrade the agreement with the observed
extinction law in the optical--UV region. However a reasonable compromise
can be found. For graphite grains we adopted the following size
distribution:
\begin{equation} 
\frac{dn_i}{da}
= \left\{ 
\begin{array}{ll} A_{i} \, n_{H} \, a^{\beta_1}  & \textrm{if
$a_b<a<a_{\rm max}$} \: ,\\ 
A_{i} \, n_{H} \, a_b^{\beta_1-\beta_2} a^{\beta_2}&
\textrm{if  $a_{\rm min}<a<a_b$} \: ,
\end{array} 
\right. 
\end{equation} 
with $a_{\rm min}=8$ \AA, $a_{\rm max}=0.25$ $\mu$m, $a_{b}=50$ \AA,
$\beta_1=-3.5$, $\beta_2=-4.0$ and $A_g=10^{-25.22}$ cm$^{2.5}/$H. As for
silicate grains, we maintained the same size distribution as DL. As a
result, 282 atoms per million H (ppM hereafter) of C are locked in dust
grains. 

In the code the size distributions have been discretized in 20 logarithmic
bins following the prescriptions given by Draine \& Malhotra (1993). Once
the bathing radiation field is specified, the emissivity of grains with
radius $a>100$ \AA\ is computed assuming that they achieve thermal
equilibrium, so that all grains of a given composition and radius emit as
gray bodies at a single temperature. Below this limit a temperature
distribution for each size bin and composition is computed following
Guhathakurta \& Draine (1989), and then the emissivity is obtained by
integrating over this distribution.

\nocite{PLB85,DrA85,DAF97,DrM93,GuD89}

\begin{figure}
\epsscale{1.0}
\plotone{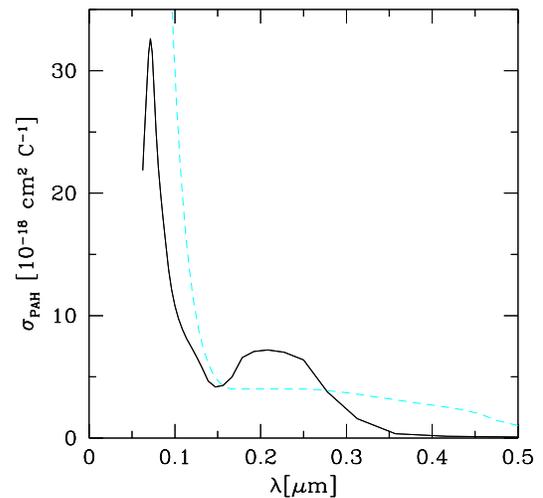}
\caption{Solid line: cross--section of PAH per carbon atom,
derived from laboratory measurements taken from L\'eger et al.\ (1989);
dashed line: the analytical law assumed by D\'esert et al.\ (1990) for
molecule with $N_C=50$ C--atoms ($N_C$ controls the
cut--off above 0.3 $\mu$m).
}
\label{figpah}
\end{figure}

To model the observed IR emission occurring in bands at 3.3, 6.2,
7.7, 8.6 and 11.3 $\mu$m we include also a population of planar PAHs.
The optical--UV absorption cross--section $\sigma_{PAH}$ is
reported in Fig.\ \ref{figpah}. This has been obtained by averaging the
cross sections of 6 different PAH mixtures measured by L\'eger et al.\
(1989) above 1200 \AA, and smoothly joining the resulting mean curve to
the coronene cross section below this limit. The same cross section is
used to take into account the effects of PAH on the extinction curve,
assuming a vanishing albedo (Fig.\ \ref{figesci}). D\'esert, Boulanger, \&
Puget (1990) have proposed an analytical description of $\sigma_{PAH}$
derived from the observed interstellar extinction curve in the
1200--3300 \AA\ range, under the assumption that these particles
dominate its EUV rise. On the other hand the extreme UV (EUV) rise can
be ascribed to small grains. Actually using DL optical efficiencies, the
EUV part of the extinction curve is produced by small grains. Our
$\sigma_{PAH}$ rests on laboratory measurements and predicts, in the
local interstellar radiation field, a PAH absorption and subsequent IR
emission lower by a factor $\sim 0.5$ with respect to the D\'esert et al.\
(1990) analytical approximation.

In order to compute the IR emission of PAHs, their heat capacity
$C_{PAH}(T)$ must be specified. A numerically convenient and accurate
enough representation of the estimate given by L\'eger, d'Hendecourt, \& De
Fourneau (1989) is:
\begin{equation}
\frac{C_{PAH}(T)}{C_{max}} = 
\left\{ \begin{array}{ll}
9.25 \times 10^{-4} \,  T   & \textrm{if $T < 800$ K} \: ,\\
2 \times 10^{-4} \, T+ .58  & \textrm{if  $800\le  T < 2100$ K} \: ,\\
1                    & \textrm{if  $T \ge 2100$ K} \: ,
\end{array}
\right. 
\end{equation}
where $C_{max}= 3 [N_t-2] k$ and $N_t=N_C+N_H$ is the total number of atoms
(carbon and hydrogen) in the molecule.  We adopt a population of PAHs with
a distribution $dn/dN_C \propto N_C^{-2.25}$ from $N_C=20$ to $N_C=280$.
Smaller molecules are easily destroyed by UV photons (Omont 1986). This
distribution is quite similar to those adopted by other authors (Dwek et
al.\ 1997, D\'esert et al.\ 1990). Astrophysical PAHs are thought to be
partially {\it dehydrogenated}: probably due to the large UV flux in the
emission regions, some of the CH bonds, responsible for the 3.3, 8.6 and
11.3 $\mu$m bands, are broken. The number of H atoms in a molecule is thus
written as $N_H = x_H N_{s,H}$, where $N_{s,H}$ is the number of hydrogen
sites and $x_H$ is the H coverage. The relationship between
$N_{s,H}$ and $N_C$ depends on the arrangements of hexagonal cycles in the
molecule. The ratio $N_C/N_{s,H}$ is rather constant around 2 for
laboratory PAHs molecules, for which $N_C \lesssim 50$. However, the
structure of typical interstellar PAHs is likely closer to that of {\it
catacondensed} PAHs (Omont 1986), which are the most compact and stable
ones and have the general formula $C_{6p^2}H_{6p}$, i.e.\ $N_{s,H}=(6 \,
N_C)^{0.5}$. For these quasi--circular molecules, usually assumed to
represent the gross features of interstellar PAHs, the radius is given by
$a=0.9 \, \sqrt{N_C}$ \AA\ if $N_C >>1$. With the above relationship
between $N_{s,H}$ and $N_C$ the typical flux ratios observed in the ISM
between CH and CC bands (Mattila et al.\ 1996) is fairly well reproduced by
our PAH model setting $x_H=0.2$

\nocite{LVH89,DBP90,LDD89}

PAHs emissivity was then computed following substantially the guidelines
given by Xu \& De Zotti (1989)\footnote{Apart two imprecisions in  their
equations: the factor 2 before $\sigma_{PAH}$ in their equation (13) is
wrong since their cross section already takes into account the two molecule
surfaces. On the other hand the integration over solid angle yields a
factor $4 \pi$ before $I_{\nu'}$ in the same equation, which in conclusion
should be multiplied by a factor $2 \pi$ on the rhs}. The adopted
abundances of PAH molecules in the diffuse ISM and in the MCs implies that
18 and 1.8 ppM of C are locked in this component respectively. Indeed there
are indications showing that in denser environments and/or in stronger UV
radiation field the relative number of small particles is significantly
diminished (Puget \& L\'eger 1989; Kim \& Martin 1996 and references
therein). In particular Xu \& De Zotti (1989) concluded that PAH molecules
are less abundant by a factor $\sim 10$ in our galaxy star forming regions
than in the diffuse gas.

As apparent from Fig.\ \ref{figesci}, this model reproduces reasonably well
both the  extinction from IR to UV and the whole cirrus emission. The model
predicts an extinction below a few observational estimates at $\lambda
\gtrsim 300 \mu$m by a factor $\sim 10$. To account for this,
Rowan--Robinson (1986, 1992) introduced {\it ad hoc} modifications of the
grain optical properties in his discretized models. We avoid in general this
approach (but see Sec. \ref{seca220}) mainly because these data have large
uncertainties (DL), as it is apparent from their scatter.
Moreover they refer to dusty environments for many respects different from
the diffuse dust, which is instead the dominant contributor to the emission
at $\lambda \gtrsim$ 100 $\mu$m in our galaxy models. Also it has been
suggested that the silicate grain absorption in the sub--mm wavelength range
may decline less steeply than $\lambda^{-2}$ (Agladze et al.\ 1996). 

The adopted mixture, even taking into account the contribution of PAH
emission features, tends to underpredict the MIR cirrus emission, in
particular the shorter $\lambda$ data. A larger quantity of small grains
and/or PAH molecules would improve the match with DIRBE observations, but
would also produce a large disagreement with observed UV--extinction (Dwek
et al.\ 1997). We prefer to adopt a compromise more balanced toward the
extinction law, since in our models and in most interesting cases the MIR
emission is in general dominated by warm dust in MCs rather than thermally
fluctuating grains in the cirrus, whilst a good specification of the
extinction properties is crucial. Moreover, DIRBE cirrus data at short
wavelength could be affected by significant systematic effects (Dwek et al.\
1997).

\nocite{PuL89,XuD89,KiM96,Row86,Row92,ASJ96}

The total (i.e.\ cirrus + MCs) dust content of a galaxy ISM depends on the
residual gas and on dust--to--gas ratio $\delta$. The residual gas mass is
provided at each time step by the chemical evolution model (see above).
Determinations of $\delta$ in our Galaxy ISM and in nearby objects range
typically from 1/100 to 1/400. For the purpose of this paper, where the
model is compared only to galaxies in the local universe, we simply set
$\delta=1/110$, the standard value of DL model. However the code is thought
to interpret data on objects in very different evolutionary stages, thus we
need in general a recipe to scale $\delta$ when the chemical conditions in
the ISM where very different. We adopt the simplest assumption $\delta
\propto Z$, with the proportionality constant adjusted to have
$\delta=1/110$ for $Z=Z_\odot$.

A recent difficulty for models relying on populations of carbon--based
grains, the so--called {\it carbon crisis}, is connected with the quite
uncertain determination of cosmic abundance of  this element. Indeed these
models were developed under the hypothesis that the measured solar system
abundances (either in meteorites or solar photosphere) were  also typical
for the ISM. For instance, DL as well as our model, requires 282 ppM of C
in grains, assuming a graphite density of 2.26 gr cm$^{-3}$. This is 55 \%
of the solar abundance which was adopted by DL as cosmic reference
abundance (but $\sim$ 80 \% of the more recent estimate in the solar
system $\simeq 365$ ppM quoted by Anders \& Grevesse 1989). However in the
last few years evidence has been accumulated that solar abundances may be
not representative of the ISM. In particular in young star photospheres
carbon is found to be only $\simeq 200-250$ ppM (e.g.\ Snow \& Witt 1995,
1996; Gies \& Lambert 1992). This, if really representative of the ISM
abundance, would put restrictions difficult to meet by carbon--based dust
models (Sofia, Cardelli, \& Savage 1994; Cardelli et al.\ 1996; Mathis
1996), especially when taking into account that about 140 ppM seem to be
in gas phase (Cardelli et al.\ 1996), leaving for dust only about 1/3 of
the C required by the DL model.

Possible ways out have been suggested: for instance big grains could be
porous, likely because they are built up by sticking of smaller ones (Mathis
\& Whiffen 1989; Mathis 1996), or dust could be to some extent prevented to
be incorporated into stars during their formation (Snow \& Witt 1996). Also
the introduction of carbonaceous grains in a form other than graphitic
(amorphous carbon, possibly hydrogenated, or composite grains), the
use of more complex size distribution than a simple power law, and an
elongated grain shape tend to alleviate the C requirements (Mathis 1996; Kim
\& Martin, 1996).

\nocite{AnG89,SnW95,SnW96,GiL92,SCS94,CMJ96,Mat96,MaW89}

Since these problems are still far from being clarified and our
major concern is to obtain a good description of the absorption and
emission properties of dust, we maintain by now a DL--type model,
looking only for the minimum modifications required to obtain an
acceptable overall match of the extinction curve from the UV to the
far--IR and of the galactic cirrus emission. It is worth noticing
that in the future this goal should be achieved with a more
efficient use of the available cosmic carbon. The main effect of
this on our spectrophotometric galaxy models will be a possible
decrease of the required dust--to--gas ratio by a factor up to $\sim
2$.

\subsection{The radiative transfer in dusty media} 
\label{secrad}

As already remarked, to get the SED from 0.1 to 1000 $\mu$m, we have to
solve the transfer equation for the radiation in presence of dust in the
case of molecular clouds and diffuse ISM.

\subsubsection{Radiation transfer in MCs and emerging spectra}
\label{secmc}

In the Galaxy virtually all the star formation activity resides in molecular
clouds. Maps of CO and other tracers, show that MCs are non-uniform, highly
structured objects, containing density-enhanced regions, the cores, wherein
star formation actually occurs. This implies the clustering of young stars
at different locations inside giant molecular clouds, as confirmed by
infrared imaging. The first evolutionary stages are hidden at optical
wavelengths and a significant fraction, if not all, of the energy of young
stellar objects is reprocessed by dust and radiated in the IR, only a minor
fraction being reradiated as recombination lines. The powerful stellar winds
and outflows, and the ionizing flux from massive stars, all contribute to
the destruction of the molecular clouds in a time scale comparable to the
lifetime of OB stars, $\sim 10^{6}-10^{7}$ yr. Thus stars gradually get rid
of their parent gas and become visible at optical wavelengths. The typical
dust densities in these objects are so high that even IR photons are self
absorbed, thus to compute the emitted SED the radiative transfer equation
must be solved.

The complex evolution depicted above is simulated as follows. 
A fraction $f_\mc$ of the model gas mass $M_{\rm gas}$ at $t_G$ is ascribed
to the dense phase under discussion in this section. Recent estimates in our
Galaxy suggest that half of the hydrogen mass is molecular H$_2$, mainly in
clouds with diameters $\gtrsim$ 10 pc. The molecular gas $M_\mc=f_\mc \,
M_{\rm gas}$ is then sub-divided into spherical clouds of assigned mass and
radius, $m_\mc$ and $r_\mc$, which may range in the typical observed
intervals $\sim 10^{5}-10^{6}\; M_{\odot}$ and $\sim 10-50$ pc respectively.
It is then supposed that each generation of stars, represented in our scheme
by a SSP, is born within the cloud and progressively escapes it. This is
mimicked by linearly decreasing the fraction $f$ of SSP energy radiated
inside the cloud with its age $t$:
\begin{equation}
f = 
\left\{ \begin{array}{ll}
1 & \textrm{if $t\le t_o$} \: ,\\
2 - t/t_o & \textrm{if  $t_o < t \le 2\, t_o$} \: ,\\
0 & \textrm{if $t > 2\, t_o$} \: .
\end{array}
\right. 
\label{equf}
\end{equation}
The model parameter $t_o$ introduced by this relation sets the fraction of
starlight that can escape the starbursting region.  
The starlight locked up inside the cloud is approximated as a single central
source. The cloud optical depth is fixed by the cloud mass and radius and by
the dust-to-gas ratio. The emerging spectrum is obtained by solving the
radiative transfer through the cloud with the code described by Granato \&
Danese (1994).

The distribution of stars in real GMCs implies a dust temperature
distribution with many hot spots and cooler regions randomly distributed. A
complete discussion of the effects of different approximations in this
complex geometrical situation, including the single central source adopted
here, can be found in Kr\"ugel \& Siebenmorgen (1994, see in particular their
Fig.\ 1a). An obvious drawback of our approach is an overestimate of the
amount of very hot dust around the source, with respect to the more
realistic approximation in which the  stars are split in many sources at
different locations. On the other hand the approximation of a central point
source is not much different from the IR spectrum predicted by a full
treatment with hot spots, if the emission from the whole cloud is considered
(Kr\"ugel \& Siebenmorgen 1994).

Since the treatment of many hot spots in the cloud would introduce many
other geometrical parameters and would slow down considerably our code due
to the loss of symmetry, we simply treat the maximum temperature $T_s$ of
the grains at the inner edge as a parameter, summarizing in some way the
geometrical parameters which would result in a lower 'true' maximum
temperature: lowering the grain 'sublimation' temperature produces a lower
average temperature with a less opaque cloud. This brings the overall
spectrum very close to that predicted with a proper treatment of the hot
spots. It is anyway interesting to note that Gordon et al.\ (1997) found
that their {\it shell geometry}, similar to that adopted here, is suited to
explain the observed optical and UV properties of starburst galaxies, while
the {\it dusty geometry}, in which dust and stars share the same spatial
distribution, is not, even allowing for clumpiness.

\begin{figure}
\epsscale{1.0}
\plotone{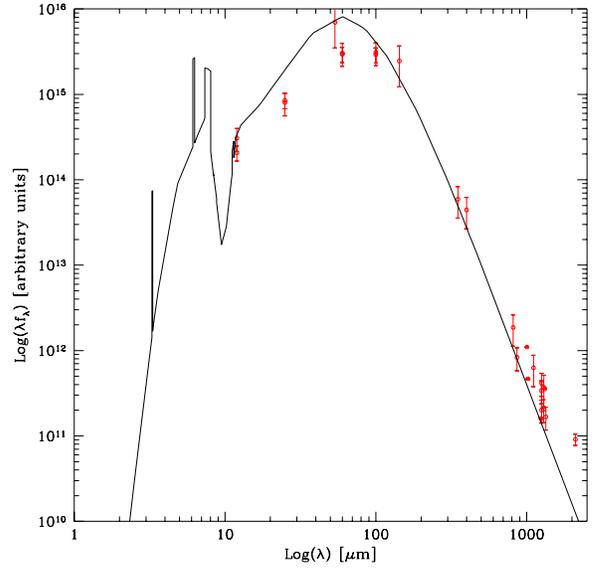}
\caption{The observed SED of W49A (Ward-Thompson \& Robson 1990), a
huge star--forming region composed of H II regions and GMCs,  is
compared with the spectrum of the MC component we use in the fit of
the overall SED of Arp 220.}
\label{figw49a}
\end{figure}

For typical values of the relevant parameters our model predicts IR
spectra of star forming regions peaking around 40--60 $\mu$m (in a $\nu
L_{\nu}$ plot), depending on the optical depth, with a rather steep
decrease at longer wavelengths. We found that this behavior reproduces
quite well the typical range of IR spectra of galactic YSOs, H II, and
star forming regions (e.g.\ Rowan--Robinson 1979; Ward-Thompson \& Robson
1990; Men'shchikov \& Henning 1996). For instance in Fig.\ \ref{figw49a}
the observed SED of W49A, a huge star--forming region composed of H II
regions and giant molecular clouds,  is compared with the spectrum of the
MC component we use in the fit of the overall SED of Arp 220. Indeed we
think that this ultraluminous infrared galaxy is a good benchmark for our
MCs model, having an IR spectrum dominated by this component (see Section
\ref{seca220}).

\nocite{Row79,WaR90,MeH96}

\subsubsection{Propagation in the diffuse ISM and cirrus emission}
\label{secdif}

Before escaping the galaxy the light arising from stellar
populations and from molecular clouds interacts with the 
diffuse dust component. 

\begin{figure}
\epsscale{1.0}
\plotone{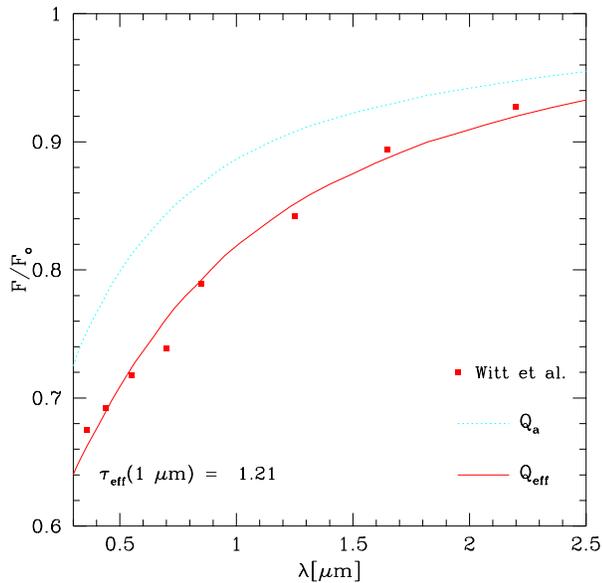}
\caption{$F/F_o$ is the ratio between the dust extinguished flux
and that expected if dust were not present. The points have
been computed by Witt et al.\ (1992) with a Montecarlo radiative
transfer code including anisotropic scattering, the dashed line
represents the result of our code taking into account only
absorption, whilst the solid line includes scattering with the
effective optical depth $\tau_{eff}^2=\tau_a (\tau_a + \tau_s)$.
The adopted geometry is that defined by Witt et al.\ as 
{\it elliptical galaxy.}}
\label{figwit}
\end{figure}

We adopt a simplified treatment of radiative transfer in the
diffuse gas which ignores dust self--absorption and approximates the
effects of optical--UV scattering by means of an effective optical
depth, given by the geometrical mean of the absorption and
scattering efficiencies $\tau_{eff}^2=\tau_a (\tau_a + \tau_s)$
(Rybicky \& Lightman 1979, p. 36). Indeed the relatively low opacity
of dust in the IR regime, where the dust emission occurs, implies
that in most cases of interest the diffuse ISM is transparent to its
own photons. Our approximation for combined scattering and
absorption processes is rigorously applicable only to an infinite
homogeneous medium and isotropic scattering. However we checked, by
comparing our results with those obtained by Witt et al.\ (1992) by
means of a Montecarlo radiative transfer code including anisotropic
scattering, that it is fairly good in most "real--world" geometrical
arrangements (see Fig. \ref{figwit})

The galaxy is subdivided into small volume elements $V_i$. The local
(angle averaged) radiation field in the $i$--th element due to the
extinguished emissions of free stars and MCs from all the elements
is computed from:
\begin{equation}
J_{\lambda,i} = \sum_k \, \frac{V_k \, (j^\mc_{\lambda,k}+j^*_{\lambda,k})
\, \exp \left(-\tau_{eff,\lambda}(i,k)\right)}{r^2(i,k)} \: ,
\end{equation}
where $\tau_{eff,\lambda}(i,k)$ and $r(i,k)$ are the effective optical
thickness and the distance between the elements $i$ and $k$ respectively.
Then the local dust emissivity is calculated as described in section
\ref{secdus}. Finally the specific flux measured by an external observer
in a given direction $\theta$ is derived as a sum over the galaxy of the
extinguished emissivity of free stars, MCs, and diffuse dust:
\begin{equation}
F_{\lambda}(\theta) = 4\pi  \sum_k \, V_k \,
j_{\lambda,k} \, 
\exp \left(-\tau_{eff,\lambda}(k,\theta)\right) \: ,
\end{equation}
where $\tau_{eff,\lambda}(k,\theta)$ is the optical thickness from
the element $k$ to the outskirts of the galaxy along the direction
$\theta$.

In the code the spheroidal systems are assumed to extend up to the
tidal radius $r_t=10^{2.2} r_c$ (of the star component), while
exponential disks are truncated at $6 R_d$, where $R_d$ is the
largest of the three components. Since more than 98$\%$ of the mass
is included within these radii, adoption of larger cut off radii
would not produce any significant difference. Deviations from the
energy balance between dust emission and absorption, which mainly
depend on the number of volume elements in which the galaxy is
subdivided, are kept within $<2-5 \%$.

\nocite{KyB87,ICG95,BST84,RyL79}

\section{Comparison with the data}

\label{seccompa}

A few examples, showing the capability of our model to reproduce the data on
starbursting and normal galaxies in the local universe, are discussed below.
We restrict ourselves to galaxies for which available photometric
observations allow a precise evaluation of the SED for UV to far--IR. The
fitting parameters are reported in Tables \ref{tabchi} and \ref{tabgeo} for
the chemical model (step 1)  and for the photometric model (step 2)
respectively. Table \ref{tabqua} summarizes a few relevant quantities derived
from those parameters. Since the purpose of this paper is to present a
procedure to take into account the radiative effects of the ISM, the
parameters $t_{inf}$ and $\nu$ (Table \ref{tabchi}), which rule the SF
history of the galaxy, where set following the general results of papers
dealing with the chemical evolution of different types of galaxies (see
Matteucci 1996 for a review), under the major constraint to get a suitable
amount of residual gas. The following parameters have been fixed to
reasonable values: galactic age $t_G=13$ Gyr; dust--to--gas mass ratio
$\delta=9 \times 10^{-3}$; exponent of Schimdt SFR $k=1$; maximum temperature
of dust in MCs $T_s=400$ K, mass of single MC $m_\mc=10^6 M_\odot$. Moreover
the mass limits of the Salpeter IMF are the standard $M_{up}=100 M_\odot$ and
$M_{low}=0.1$, but for starburst galaxies where the estimates of dynamical
mass require $M_{low}\gtrsim 0.2$, as discussed below. We also remind that,
in order to keep the number of free parameters to the minimum required by
present quality data, we set the scale lengths of stars and gas distributions
in spirals  to the same values, i.e. $R_d^*=R_d^c$ and $z_d^*=z_d^c$.

As previously noticed, the SED of the MCs component is set mainly by its
optical depth (Table \ref{tabqua}) $\propto m_\mc/r_\mc^2$, which is in some
sense the true "fitting parameter" for this component (together with $t_o$),
rather than $r_\mc$ reported in Table \ref{tabgeo}. In other words the values
found for $r_\mc$ depend on having set $m_\mc=10^6 M_\odot$, as typical for
Giant Molecular Clouds in the Galaxy, and $r_\mc$ could well be varied by a
factor $\sim 2$ producing almost equivalent fits, provided $m_\mc$ would be
adjusted to keep $m_\mc/r_\mc^2$ constant.

\subsection{Starburst galaxies}

\subsubsection*{M82}

\begin{figure}
\epsscale{1.0}
\plotone{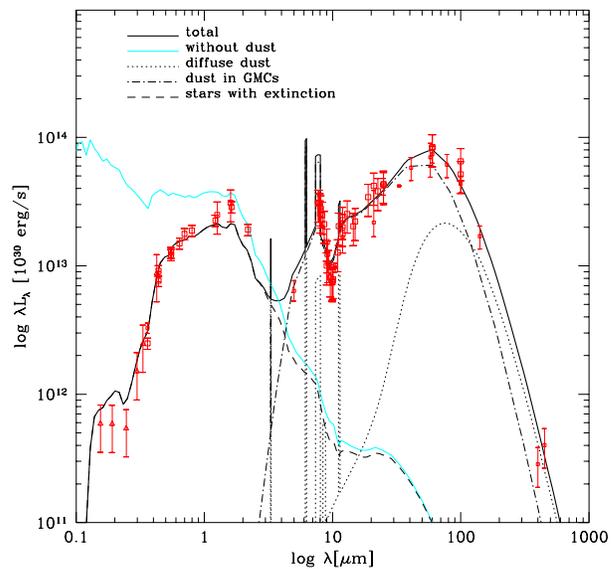}
\caption{
Fit to the SED of M82. Data are from Code \& Welch (1982), Soifer et al.\
(1987), Klein et al.\ (1988),Cohen \& Volk (1989), Van Driel et al.\ (1993),
Ichicawa et al.\ (1994,1995).}
\label{figm82}
\end{figure}

\begin{figure}
\epsscale{1.0}
\plotone{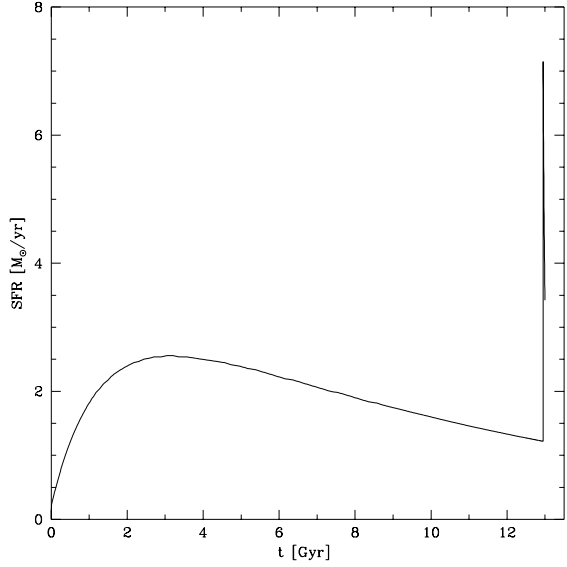}
\caption{
M82: SFR as a function of galactic time.}
\label{figm82sfr}
\end{figure}

In the prototype starburst galaxy M82 the burst was probably triggered by
the interaction with M81, some $10^8$ yr ago (Solinger, Morrison, \&
Markert 1977). Thanks to the proximity of this system ($D = 3.25$ Mpc) a
wealth of data do exist, providing a well sampled full coverage of the SED
at different angular resolutions, as well as other observational
constraints. The fit (Fig.\ \ref{figm82}) is obtained by evolving for 13
Gyr an open system with a final baryonic mass of $1.8 \times 10^{10} \,
M_\odot$, a factor 6 less than the estimated total dynamical mass (Doane
\& Mathews 1993). We adopted $M_l = 0.2 \, M_\odot$ as lower limit of the
Salpeter IMF. A lower value would require a baryonic mass closer to the
dynamical mass. The assumed parameters provide a SFR raising from 0 to
about 3 $M_\odot/\rm{yr}$ in the first 3 Gyr, then smoothly declining to
1.35 $M_\odot/\rm{yr}$ at $t=13$ Gyr. To this gentle star formation
history, which leaves a gas fraction of 0.064, we have superposed an
exponential burst processing 18\%  of the residual gas in the last $5
\times 10^7$ yr, with an e--folding time also of $5 \times 10^7$ yr
(Figure \ref{figm82sfr}). Note that our results are almost independent on
the precise evolution of the SFR in the burst: even a constant burst
yields very similar estimates of the relevant quantities. Also the
duration of the burst could be doubled or halved with small adjustment of
the other parameters without damaging the quality of the SED fit. Thus we
end with a total gas mass of $8.6 \times 10^8 \, M_\odot$, $8 \%$ of which
is ascribed to the molecular component, organized in clouds with
$m_\mc=10^{6} \, M_\odot$ and $r_\mc=16$ pc. These clouds reprocess almost
completely the starlight due to the burst. The system is assumed to follow
a King profile with $r_c^\star=150$ pc for stars and $r_c^c=200$ for
diffuse gas.

The masses we ascribe to the various components favorably compare with radio
estimates of total gas masses $\sim 10^9  \, M_\odot$ (Solinger et al.\
1977), with CO determinations of gas in molecular form $\sim 10^8 \,
M_\odot$ (Lo et al.\ 1987; Wild et al.\ 1992), and with an upper limit
$\lesssim 3 \times 10^8 \, M_\odot$ to the mass of stars formed in the
burst, suggested by dynamical considerations  (McLeod et al.\ 1993). Also,
in our proposed model the predicted supernovae rate (SNR) is between 0.05
and 0.1 yr$^{-1}$, depending on the adopted lower limit of stellar mass
yielding to supernovae explosion 9 or 6 $M\odot$ respectively. Observational
estimates suggest a SNR in the range 0.07--0.3 yr$^{-1}$ (McLeod et al.\
1993; Doane \& Mathews 1993). Increasing the lower limit of the IMF in the
burst to, for instance, $M_l = 1.0 \, M_\odot$ would increase the SNR by a
factor $\sim$ 2. In conclusion the model required to nicely reproduce the
observed SED turns out to be in agreement with independent estimates of
masses and of SNR.

\nocite{SMM77,DoM93,LCM87,WHE92,MRR93}

\subsubsection*{NGC 6090}

\begin{figure}
\epsscale{1.0}
\plotone{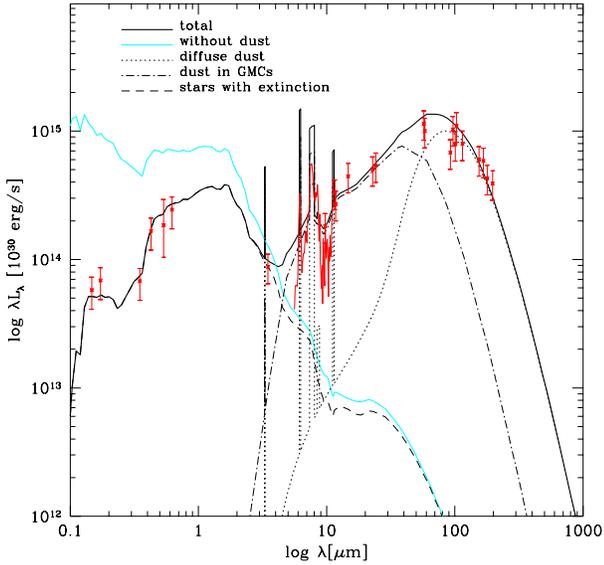}
\caption{
Fit to the SED of NGC 6090. Data are from Mazzarella \& Boroson (1993),
Acosta--Pulido et al.\ (1996), Gordon et al.\ (1997).}
\label{fign6090}
\end{figure}

This strongly interacting galaxy at 175 Mpc ($\rm{H}_o=50$ km/s/Mpc) has
been observed by ISO from 2.5 to 200 $\mu$m (Acosta-Pulido et al.\ 1996).
The SED resulting from the combination of  these data with previously
published optical photometry is nicely reproduced by our model (Fig.\
\ref{fign6090}). The differences in the parameters with respect to M82,
apart from an up-scale in involved baryonic mass ($4.1 \times 10^{11} \,
M_\odot$), are aimed at enhancing the diffuse dust emission at $\lambda
\gtrsim 100 \, \mu$m, which, as noticed by Acosta-Pulido et al., is not
reproduced by published starburst models. Thus the pre-burst SFR has been
adjusted to leave a larger gas fraction (0.127), 3.2 \% of which is
processed by the burst. The molecular star-forming clouds accounts only
for 0.5 \% of the gas left by the burst.

\subsubsection*{Arp 220} 
\label{seca220}

Arp 220 is an archetypal ultraluminous infrared galaxy (ULIRG), most likely
the result of a recent merging between two gas-rich galaxies, for which
ISOPHOT data have been now published (Klaas et al.\ 1997), and whose K--band
light profile resembles that of a typical elliptical galaxy. In this object
there is evidence of both starburst as well as Seyfert activity, but ISO
spectroscopy led to the conclusion that the IR luminosity is primarily
($\gtrsim 90$ \%) powered by starburst (Sturm et al.\ 1996). The fitting
model requires a strong burst, converting into stars $2.5 \times
10^{10} \, M_\odot$, i.e.\ as much as 11\% of the total baryonic mass, and
a high fraction 50\% of residual gas in star--forming clouds. The
gas fraction before the burst is 0.25. As a result, the IR and
sub--millimetric emission is everywhere dominated by dust associated with
star forming regions.  The mass in molecular gas is $\sim 1.6
\times 10^{10} \, M_\odot$, in agreement with CO estimates (Solomon et al.\
1997; Scoville, Yun, \& Bryant 1997). From a determination of the CO rotation
curve Scoville et al.\ (1997) infer a dynamical mass of $\sim 3-6 \times
10^{10} M_\odot$ within $r \simeq 1.5$ kpc (depending on the adopted
potential), in agreement with the mass requested by our model within the
same radius $\simeq 4.1 \times 10^{10} \, M_\odot$.

\begin{figure}
\epsscale{1.0}
\plotone{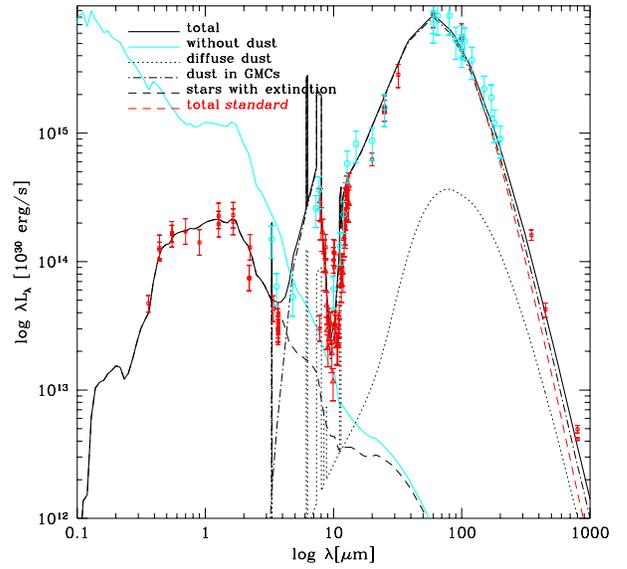}
\caption{
Arp 220: in this case the wavelength dependence of grain cross--section has been
modified above $100 \mu$m from $\propto \lambda^{-2}$ to $\propto
\lambda^{-1.6}$. The dashed line above 100 $\mu$m represents the model
prediction with standard $\lambda^{-2}$ decline. Data are from Carico et al.\
(1988), Sanders et al.\ (1988), Smith et al.\ (1989), Carico et al.\ (1990),
Wynn--Williams \& Becklin (1993), Rigopoulou et al.\
(1996), Klaas et al.\ (1997).}
\label{figa220}
\end{figure}

We find that the sub--millimetric data of Arp 220 are underpredicted
by a factor $\sim 2-3$ by any model which successfully reproduces the
remaining SED, unless the long wavelength decline  of grain absorption
efficiency is reduced, say from $\propto \lambda^{-2}$ to $\propto
\lambda^{-1.6}$ above $\simeq 100 \mu$m (see Agladze et al.\ 1996 for
laboratory measurements supporting this possibility for silicate
grains). In Fig.\ \ref{figa220} we report fits obtained adopting either
the modified as well as the standard DL cross--sections.

\nocite{SLG96,SDR97,SYB97}

\subsection{Normal galaxies}

In the following we examine objects that do not show clear signs of enhanced
SF or nuclear activities in the SED nor in the morphology,  namely 3
late--type spirals and a giant elliptical template. Indeed elliptical
galaxies form a relatively homogeneous class, while this is not the case for
spirals. We fit spirals later than Sb, which are clearly disk--dominated,
and we adopt for them an exponential geometry (see Sec.\ \ref{secgeo}).

\subsubsection*{M51}

\begin{figure}
\epsscale{1.0}
\plotone{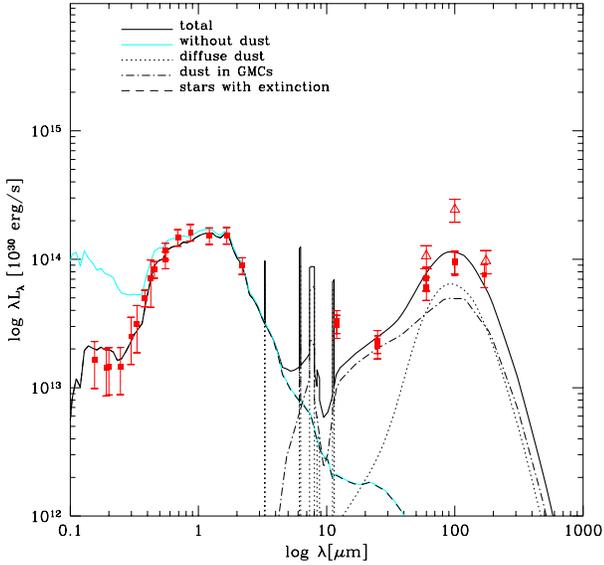}
\caption{
Fit to the SED of the Sbc galaxy M51. Data are from Buat et al.\ (1989), Evans
(1995), de Vaucouleurs et al. (1991) (RC3), Code \& Welch (1982), Young et
al.\ (1989), Rice et al.\ (1988), Devereux \& Young (1990,1992), Smith (1982),
Hippelein et al.\ (1996) (ISO, triangles).}
\label{figm51}
\end{figure}

In Fig.\ \ref{figm51} we present a fit to the SED of the nearly face--on
($i=20^o$, Tully 1974) Sbc galaxy M51 (NGC 5194), taken at a distance
$D=9.6$ Mpc (Sandage \& Tammann 1975). Since the discrepancy between ISO
and IRAS data at 60 and 100 $\mu$m could be due to an excess of the ISO
point spread function (Hippelein et al.\ 1996), we fit the IRAS data. The
model has a baryonic mass of $1.55 \times 10^{11} \, M_\odot$, whose
evolution leaves at 13 Gyr a gas fraction of 0.067, 70\% of which in
molecular form. The mass of ISM in the molecular and diffuse components, as
well as their relative fraction is in agreement, within a factor $\le 2$,
with estimates by Scoville \& Young (1983), Young et al.\ (1989), and
Devereux \& Young (1990). Gas and stars are exponentially distributed with
the same scale--lengths, 4.7 and 0.4 kpc for the radial and vertical scales
respectively.  The former value is in good agreement with those estimated by
Beckman et al.\ (1996) from observed brightness distributions in different
optical bands. Due to the low energy output observed at $\lambda \lesssim
0.2 \mu \rm{m}$, stars born during the last $5 \times 10^6$ yr are hidden
inside molecular clouds.

The striking correspondence between the 15 $\mu$m emission mapped by ISOCAM
and the H$_\alpha$ emission indicates that the MIR is powered by recent
star--formation (Sauvage et al.\ 1996). As for the far--IR (FIR),
somewhat contradictory claims have been reported: Devereux \& Young (1992),
comparing the radial distributions of FIR, H$_\alpha$, H I, and H$_2$
emission, concluded that the same holds true in the range 40--1000 $\mu$m,
while Hippelein et al.\ (1996) found no obvious correlation between FIR
ISOPHOT maps and H$_\alpha$ fluxes. This apparently complex situation is not
surprising, since according to the model the MIR emission is provided by the
MCs component, while above $\sim 60 \, \mu$m the diffuse dust gives a
comparable contribution, which however is in part powered by young stars.

\nocite{Tul74,SaT75,ScY83}

\subsubsection*{M100}

\begin{figure}
\epsscale{1.0}
\plotone{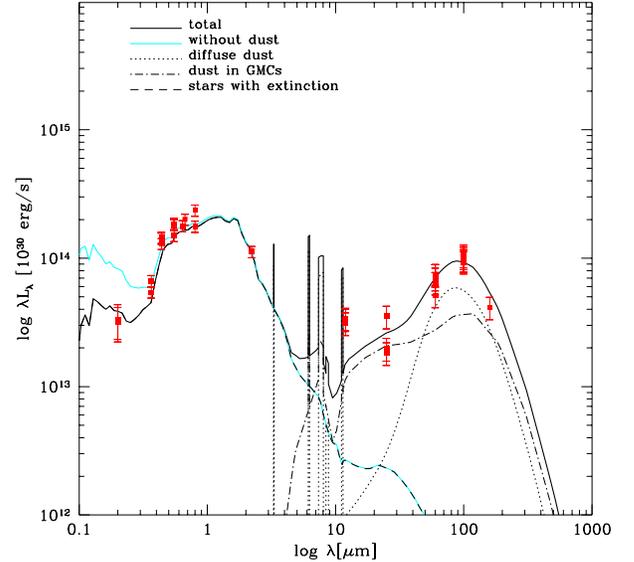}
\caption{
Fit to the SED of the Sbc galaxy M100. Data are from Buat et al.\ (1989), Donas
et al.\ (1987), De Jong et al.\ (1994), Stark et al.\ (1989), RC3, Devereux
\&Young (1990), Young et al.\ (1989), Helou et al.\ (1988), Knapp et al.\ (1987).}
\label{figm100}
\end{figure}

The Sbc galaxy M100 (NGC 4321) is the largest spiral in the Virgo Cluster.
We adopt a distance $D=20$ Mpc and an inclination angle $i=30^o$. The model
in Fig.\ \ref{figm100} has a baryonic mass of $2 \times 10^{11} \, M_\odot$,
with a residual gas fraction of 0.048, $80 \%$ in molecular form. Indeed,
according to the gas masses estimated by Young et al.\ (1989) and Devereux
\& Young (1990), larger than ours by a factor $\simeq 2$, the ISM seems to
be dominated by the molecular component. The location of M100 in the central
regions of the Virgo Cluster could affect some of its properties, as the H I
distribution that shows a sharp edge in correspondence to the optical radius
(Knapen et al.\ 1993). This could be the reason for the warm FIR SED
observed in M100, as compared to M51 and NGC 6946, despite their
morphological similarity. The steep decline of the spectrum in the sub--mm
region has been interpreted by Stark et al.\ (1989) as indicating that the
emitting grains are warm and small. In our model this is not required since
the fit is obtained with standard diffuse dust. Stars and dust share the
same radial and vertical scale--lengths, 5 and 0.4 kpc respectively, in
agreement with the values derived by Beckman et al.\ (1996) for the stellar
component and with the H I distribution.

\nocite{KCB93}

\subsubsection*{NGC 6946}

\begin{figure}
\epsscale{1.0}
\plotone{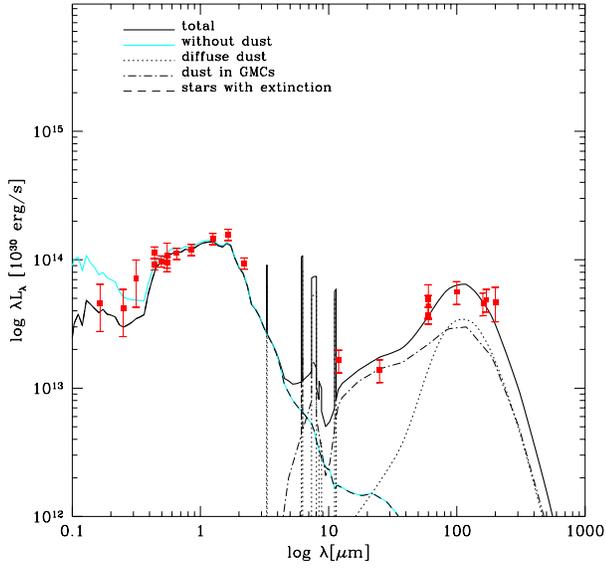}
\caption{
Fit to SED of the Scd galaxy NGC 6946. Data are from Rifatto et al.\ (1995),
RC3, Engargiola (1991), Devereux \& Young (1993), Rice et al.\ (1988), Tuffs et
al.\ (1996).}
\label{fign6946}
\end{figure}

The fit to this Scd galaxy at distance $D=6.72$ Mpc (Rice et al.\ 1988) and
inclination angle $i=34^o$ (Considere \& Athanassoula 1988), is shown in
Fig.\ \ref{fign6946}. The total gas mass in the model, $1.04 \times 10^{10}
\, M_\odot$ i.e.\ 8.3\% of the baryonic mass,  agrees with the molecular
plus neutral hydrogen mass given by Young et al. (1989) or Devereux \& Young
(1990) (reported at 6.72 Mpc), while the fraction we ascribe to the
molecular component is a factor $\simeq 2$ higher.

The contribution of young and old stellar populations to dust heating in this
object has been considered by several authors. Devereux \& Young (1993),
comparing the radial distributions of FIR, H$_\alpha$, H I, and H$_2$,
conclude that the 40 to 1000 $\mu$m luminosity is dominated by dust
associated with molecular gas heated by young stars. However ISO data
reported by Tuffs et al.\ (1996) reveal an extended FIR emission out to a
radius of $8'$, with a scale--length similar to the R--band one, while little
H$_\alpha$ emission has been detected beyond $r \sim 6'$. This suggests that
part of the observed cold FIR SED of NGC 6946 could be due to dust associated
with a diffuse H I gas, which indeed extends out to $r \sim 15'$ (Boulanger
\& Viallefond 1992), and heated mainly by old stellar populations. Malhotra
et al.\ (1996) find that the radial scale--lengths at 7 and 15 $\mu$m are
similar to those in H$_\alpha$ and H$_2$ but much shorter than those in
R--band and H I, consistent with a warm dust emission primarily heated by
young massive stars. In the proposed fit the MIR emission arises from MCs
dust heated by newly born stars. The SED above $\sim 60 \, \mu$m is almost
equally contributed by MCs and diffuse dust, but, since $t_o=2.5$ Myr, even
the latter is predominantly heated by young stars.

Tacconi \& Young (1986) estimate a scale--length of 9.6 kpc for the cold ISM
component, whilst for the star distribution their values range from 4 to 8
kpc, depending on the band and on the galactic component (disc or arms)
considered. The model scale--lengths are set to 8 for both stars and the
diffuse gas.  The vertical scale--lengths are 1 kpc to keep the diffuse dust
emission sufficiently cold.

\nocite{CoA88,BoV92,MHV96,TaY86}

\subsubsection*{Giant Ellipticals} \label{secgiaell} 

\begin{figure}
\epsscale{1.0}
\plotone{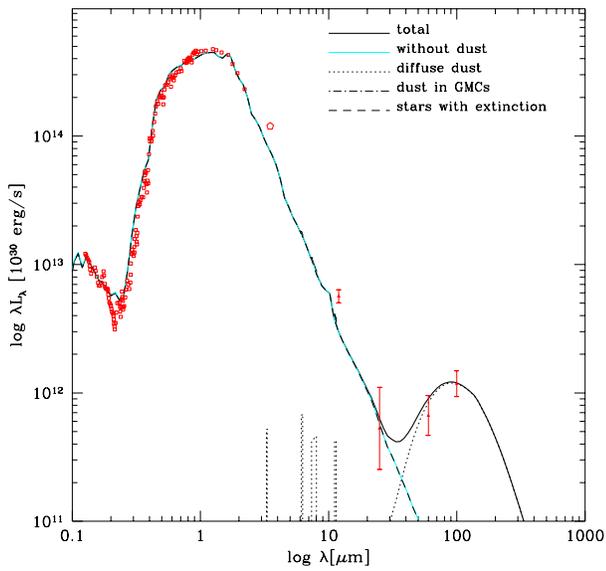}
\caption{
A fit to a template SED of a giant elliptical galaxy.}
\label{figeli}
\end{figure}

Ellipticals constitute a class of objects with fairly homogeneous spectral
properties. It is then interesting to test our model against an average
SED of giant E galaxies, constructed combining the Arimoto's template
(1996) from 0.12 to 2.2 $\mu$m with the median IRAS over B band fluxes of
bright ellipticals estimated by Mazzei et al.\ (1994) and the average
(K-L) color computed by Impey et al.\ (1986) (Fig.\ \ref{figeli}). The fit
has been obtained with a `classical' star formation history model for
ellipticals: an open model with infall and galactic wind. The high
efficiency $\nu = 2$ causes a huge SFR in the first 1.2 Gyr interrupted by
the onset of the galactic wind (see Tantalo et al.\ 1996 for a discussion
of these values). The object is observed at 13 Gyr, when a comparatively
very small fraction of galactic mass in a diffuse dusty component is
sufficient to produce the observed IRAS emission. Note that this gas is
not provided by the chemical model, according to which the bulk of the ISM
is ejected by the galactic wind which stops the star formation. Thus in
this case the mass in the diffuse dust $1.5 \times 10^7 \, M_\odot$ is a
fitting parameter of our photometric model. This dusty ISM could arise
from evolved stars of the passively evolving galaxy, by cooling flows or
by merging activity. It is also worth noticing that $r_c$ needs to be much
greater for the ISM than for stars (6 and 0.4 kpc respectively): for
smaller values of $r_c^c$ the cirrus emission would be too warm because
dust would be more concentrated in the central regions where the radiation
field is higher. Other authors already suggested that diffuse dust in
elliptical galaxies needs to be less concentrated than stars, on the basis
of either optical color gradients (e.g.\ Wise \& Silva 1996) or IRAS
colors (e.g.\ Tsai \& Mathews 1996). As discussed in section \ref{secdif},
an alternative proposed way to describe this lower concentration is to
adopt the same $r_c$ for both stars and diffuse dust, decreasing the
exponent $\gamma=1.5$ in the King law of stars. However for this SED,
adjusting $\gamma^c$ and $r_c$ with the constrain $r_c^c=r_c^*$, we found
only marginally acceptable fits in the IRAS regime, significantly worse
than that presented in Fig.\ \ref{figeli}.

\nocite{SBB96,BPK96,Ari96,IWB86}

\section{Discussion}

When reproducing  starburst galaxies, our model is characterized by the
variations of 6 parameters related to the total mass and star formation
history and of 5 geometrical parameters (Tables \ref{tabchi} and
\ref{tabgeo}). For normal disc galaxies, the parameters become 3 and 5
respectively.  In order to fit the elliptical galaxy template, 7
parameters have been adjusted, including the epoch at which the SF
activity is stopped.

The masses reasonably constrained by matching the SED are the baryonic mass
(though dependent on the IMF, particularly its lower limit) mostly from the
energy emitted in the near--IR (NIR) by the old stellar component, and the mass
of dust, from its MIR--FIR emission. Therefore the mass of gas present in the
galaxy is constrained from the fit, provided $\delta$ is known. Our assumption
$\delta=9\times 10^{-3}$ is likely the best guess for our Galaxy, however in
different systems and in the Galaxy itself, variations of a factor of a few are
commonly quoted, e.g.\ Rand, Kulkarni, \& Rice (1992) for M51 or Tuffs et al.\
(1996) for NGC 6946. As a consequence, our estimated gas masses suffer by a
similar uncertainty. The SFR history of the old stellar component is constrained
by the spectrum and, for starburst galaxies, also by the requirement that a
certain amount of gas is available for the latest burst. The three analyzed
starbursts exhibit rather different star formation histories, with the average
SFR of the old component inversely proportional to the strength of the SFR in the
burst. Estimates of masses in stars and gas, as well as of SF and SN rates from
observations other than broad band spectra, discussed in the previous sections
and below, nicely agree with those provided by our SED fitting.

\nocite{RKR92}

\subsection{Starbursts and ULIRGs}

The  SFR in a starbursting region can be estimated on the basis of the FIR
luminosity, under the assumption that dust reradiates the overall
bolometric luminosity and after calibration with stellar synthesis models.
Kennicutt (1997) proposes $\mbox{SFR}[M_{\odot} \, \mbox{yr}^{-1}]=L_{FIR}/
(2.2 \times 10^{43} \, \mbox{ergs} \, \mbox{s}^{-1})$, with calibration
derived from models of Leitherer \& Heckman (1995). This relationship
properly applies to star forming regions, while the extension to a galaxy
as a whole is dangerous, since a non negligible portion of the FIR
luminosity may arise from cirrus emission powered also by relatively old
stars. However our model accurately determines the warm component
associated to the MCs where star formation is occurring ($L_\mc$ in Table
\ref{tabqua}). For M82 our fit predicts an overall FIR luminosity
associated to the MCs component of $1.1\times 10^{44}$ ergs/s, which
following Kennicutt translates in SFR=5.0 $M_{\odot}$ yr$^{-1}$.
Similarly, the SFRs inferred from the luminosity of the warm component is
57 $M_{\odot}$ yr$^{-1}$ and 460 $M_{\odot}$ yr$^{-1}$ for NGC 6090 and Arp
220 respectively. These figures are in very good agreement with those
derived by fitting the SEDs and reported in Table \ref{tabqua}. Thus while
Kennicutt's calibration is confirmed, we stress that this refers only to
the warm component, which contributes $\simeq 50\%$ of $L_{FIR}$ in NGC
6090. This source of uncertainty adds to the obvious effect of the adopted
IMF in the conversion from observed IR luminosity to SFRs.

Whilst the very recent (say in the last 10 Myr) SFR is relatively well
constrained by the observed warm IR emission, the burst duration, and
therefore the total mass converted into stars, can be varied within a
factor $\sim 2$, still yielding the correct $L_{FIR}$ and spectra after
readjustments of other parameters. However the supernovae rates
deduced from observations put further constraints on the average SFR
and duration of the burst. In the well studied case of M82 the model
predicts a SN rate in the interval deduced by observations.

The SFR can also be inferred from the ionizing luminosity $L_{Lyc}$, which
can be derived from recombination lines. With the adopted IMF we find
$L_{Lyc}[10^{44} \, \mbox{ergs} \, \mbox{s}^{-1}] \simeq 0.017 \, \mbox{SFR}
(M_{\odot}\, yr^{-1})$, which holds with very small deviations for the six
studied star--forming galaxies (see Table \ref{tabqua}). The uncertainties
connected to extinction corrections of observed line fluxes are minimized
employing transitions occurring in the IR regime, such as Br$\gamma$ line
($\lambda=2.17  \mu$m). Calzetti (1997) and Kennicutt (1997), using Leitherer
\& Heckmann (1995) results, find $\mbox{SFR}[M_{\odot} \mbox{yr}^{-1}]
=L(Br\gamma) / 1.6 \times 10^{39}\mbox{ergs} \, \mbox{s}^{-1}$ adopting a
Salpeter IMF within 0.1--100 $M_{\odot}$ or $\mbox{SFR}[M_{\odot}
\mbox{yr}^{-1}]=L(Br\gamma) / 3.7 \times 10^{38} \mbox{ergs} \,
\mbox{s}^{-1}$ when the  mass range is 0.1--30 $M_{\odot}$. 

The $(Br\gamma)$ line luminosity $L(Br\gamma)=9.2 \times 10^{40} \,
\mbox{ergs} \, \mbox{s}^{-1}$ of NGC 6090 (Calzetti, Kinney, \&
Storchi--Bergmann 1996) corresponds to a SFR=56 $M_{\odot}$
yr$^{-1}$ with the assumption of the 0.1--100 $M_{\odot}$ mass range, only
18$\%$ less than that used in our model, while the smaller range in the IMF
would predict a SFR higher by a factor 4. Kennicutt (1997) has shown that
there is a clear trend for SFRs derived from $L_{FIR}$ to be larger than
those estimated from the $L(Br\gamma)$, suggesting that extinction is non
negligible even at near--IR wavelengths.

\nocite{CKS96,Cal97,GLS97}

Genzel et al.\ (1997) reported the ratio of the far--IR to Lyman continuum
luminosity $L_{FIR}/L_{Lyc}$ of starbursts (including M82) and ULIRGs,
derived from near and mid--IR recombination lines. For 12 starburst galaxies
the median value is $L_{FIR}/L_{Lyc}\simeq 16$. The model ratio depends on
the adopted IMF. Our fit to M82 overall SED yields $L_{FIR}/L_{Lyc}\simeq
12$, where $L_{Lyc}$ is the unextincted luminosity of the (young) stellar
populations below 912 \AA\, in very good agreement with the value found by
Genzel et al.\ (1997). For NGC 6090 the fit predicts $L_{FIR}/L_{Lyc}\simeq
10$, using $L_\mc$ to evaluate the model $L_{FIR}$ powered by the burst,
well within the range of the values inferred by the same authors. As for Arp
220, which properly belongs to the ULIRG class, higher values $15\lesssim
L_{FIR}/L_{Lyc}\lesssim 112$ are inferred from IR recombination lines,
whereas our model predicts $L_{FIR}/L_{Lyc}\simeq 11$. The discrepancy is
likely due to the uncertain large correction for IR extinction in this
object. Actually Genzel et al.\ (1997) use a screen obscuration with
$A_V=45$, while we find $A_V\simeq 150$ in MCs of this object. Interestingly
enough, the 15 ULIRGs (including Arp 220) studied by Genzel et al.\ (1997),
exhibit a median $L_{FIR}/L_{Lyc}\simeq 40$, larger by a factor of $\sim
2.5$ than that inferred for starburst galaxies. This result can be explained
by larger obscurations or by a softer intrinsic Lyman continuum, pointing to
an IMF less rich in massive stars or to an older starburst. In the models we
used bursts began 0.05 Gyr ago, but still rather active due to the large
e--folding time $t_e=0.05$ Gyr.

It is worth noticing that our model does not account for the ionizing
radiation converted into recombination lines, which are much less absorbed
by dust than the Lyman continuum. In the present version of the code the UV
radiation can be only converted directly to IR photons through dust
absorption. However the energy budget is not significantly affected, since
the observed luminosity in recombination lines is only a few percent of the
bolometric luminosity of starburst galaxies (e.g.\ Genzel et al.\ 1997).

The observed UV emission at $\lambda\gtrsim 1000$ \AA\ is crucial in
determining the fraction of the SSPs escaped from the parent molecular
clouds. In our model the UV flux emerging from a single galaxy depends on the
age $t_o$ after which a SSP starts to get out from the molecular cloud. The
UV photons are affected also by absorption in the diffuse ISM. In the case of
M82 and Arp 220 the measured UV flux is a tiny fraction of the bolometric
luminosity (see Figg.\ \ref{figm82} and \ref{figa220}). This implies that the
stars of the burst are still inside their parent clouds. Indeed in these two
objects $t_o$ is about equal to the time since burst ignition 0.05 Gyr. By
converse for NGC 6090 the data show that the UV is not a negligible fraction
of the total, implying for our model  $t_G-t_{burst} > t_o=0.018$ Gyr. As a
result stars with lifetime $\ge 18$ Myr (corresponding to $M\lesssim 10
M_{\odot}$) are no more embedded within the clouds and contribute
significantly to the ISRF. Actually the luminosity we ascribe to the diffuse
cold component equals the warm one in NGC 6090, whereas it is 25$\%$ and
9$\%$ for M82 and Arp 220 respectively.

Calzetti, Kinney, \& Storchi--Bergmann (1994) computed the "intrinsic depth"
of the 2175 \AA\ dust absorption feature  $\eta$ (see their Eq.\ 23) in a
number of starburst galaxies, finding that $-0.15<\eta<0$. For the three
starbursts we find $\eta$ ranging from -0.15 to -0.08. Since the dust
properties we adopt follows very well those of our galaxy, at least in these
objects the relative weakness of the 2175 \AA\ feature can be ascribed to
the adopted geometry, without invoking possible variations of extinction
curve.

In conclusion for starbursts, the spectral coverage from  UV to
submillimeter wavelengths allows a robust evaluation of the luminosity of
stars involved in the bursts and of the general interstellar radiation
field.  The latter is contributed by the long-lived stellar populations
and by the stars produced in the burst but old enough to get out from the
original MCs. Our model well describes the complexity of a starburst galaxy,
and is flexible enough to reproduce the differences among them. It uses a
reasonable number of parameters, which have a well defined physical meaning
and compare favorably to values derived from observations other than broad
band spectra.

\subsection{Spirals}

The SEDs of three spiral galaxies, namely M51, M100, and NGC 6946, have been
well reproduced by our model. An interesting result is the significantly
lower $t_o$ in spirals with respect to starbursting objects. This derives
from their smaller observed ratios $L_{FIR}/L_{UV}$. In the three spirals
analyzed here, the newly born SSPs quite soon get out from the parent clouds
and significantly contribute to the heating of the diffuse dust. The expected
contribution is larger for NGC 6946 and M100 than for M51. As a consequence,
a significant portion of the far--IR luminosity emitted by the diffuse dust
(almost all for NGC 6946 and M100) should be inserted in the budget to infer
the SFR in spiral galaxies, in agreement with the findings of Devereux et
al.\ (1994) and Buat \& Xu (1996). Taking into account this, the SFRs derived
from the FIR luminosities, through the conversion suggested by Kennicutt
(1997), are in very good agreement with those used by the model.

\nocite{DPW94,BuX96,Ken97,KTC94}

The SFRs in our models, through the conversion of Kennicutt (1997)
$L(H_\alpha)[\mbox{ergs s}^{-1}] = 1.26 \times 10^{41} \times SFR[M_\odot \,
\mbox{yr}^{-1}]$, yield $H_{\alpha}$ luminosities of 7.6, 8.8, and $7.6 \,
10^{41}$ ergs s$^{-1}$ for M51, M100, and NGC 6946 respectively, to be
compared with observed values (Kennicutt, Tamblyn, \& Congdon 1994,
reported to our adopted distances) of 4.9, 5.0, and $2.2 \times 10^{41}$
ergs s$^{-1}$. These lower values would imply an internal H$_\alpha$
extinction of $A_{H\alpha}=0.5$, 0.6, and 1.3 magnitudes in the three
spirals, consistent with the average value 1.1 adopted by Kennicutt (1997),
based on a comparison of free--free radio and H$_\alpha$ fluxes.

It is also worth noticing that masses in stars, dust, and gas derived by
the model, as well as geometric parameters such as the disc scale length
are in good agreement with independent estimates. Our model includes the
main physical aspects of star formation, stellar evolution, and dust
absorption in spiral galaxies and allows a full exploitation of the
broad band data in order to explain their present status and their past
history.

\subsection{Ellipticals}

The model with a relatively small number of parameters is able to produce a
very good fit to the template SED of giant elliptical galaxies (see Fig.\
\ref{figeli}). Most excitement about spheroidal galaxies is related to their
evolution. Indeed it has been suggested that, during the initial phases of
star formation, they might look very similar to local violent starbursts
(such as Arp 220), since the chemical enrichment and, as a consequence, the
dust formation are very quick processes when SFRs are very high (Mazzei et
al.\ 1994; Franceschini et al.\ 1994). This initial phase, if confined to
high enough redshifts $z\leq 2-2.5$, is the most natural way to produce the
total energy and the shape of the Far--IR Background (Franceschini et al.\
1991; Franceschini et al.\ 1994; Burigana et al.\ 1997), which has been
tentatively detected by Puget et al.\ (1996) and recently confirmed by
Hauser et al.\ (1998) and Fixsen et al.\ (1998). Also we expect that
in the ISO extra deep surveys we may detect individual dusty spheroidal
galaxies at substantial redshifts. Our model is able to describe in detail
the evolution of elliptical galaxies in all the relevant range of
frequencies, and will be exploited to a comprehensive discussion of galaxy
counts, related statistics, and astrophysical backgrounds.

\nocite{FDT91,BDD97,PAB96,Hau98,Fix98}

\subsection{Conclusion} 

We showed that the model and the related numerical code are extremely
efficient in deriving a wealth of information from broad band spectra of
starburst and spiral galaxies. Indeed, when SEDs from the UV to the sub--mm
range are available, we are able to quantify the effects of dust
reprocessing on observations, gaining information on very substantial
quantities such as SFR, the IMF and the past history of the galaxies. The
model can be implemented in studies of local starburst and normal galaxies.
It can also be used to trace back the history of the different classes of
galaxies and, by confronting predictions and observations of number counts
and redshift distributions in different bands, to understand their
cosmological evolution. In particular we will use the model to investigate
the role of dust during the early phase of galaxy formation, exploiting
available and soon coming data on galaxy counts at wavelengths ranging from
UV to radio bands.

Updated information on the code described in this study (GRASIL) can be
found on the WEB at {\it http://asterix.pd.astro.it/homepage/gian/}.

\acknowledgments

This work was supported by MURST and ASI under contract ARS--96--86.
We are indebted with the anonymous referee for suggestions which led us
to substantial improvements to the model and its presentation.


\newpage

\nocite{SSM87,KWM88,CoV89,VDD93,IYI95,IVA94}

\clearpage

\nocite{AKL96}

\nocite{CSS88,SSE88,SAR89,CSS90,WyB93,RLR96,KHH97}

\nocite{BDD89,DDC91,CoW82,YXK89,RLS88,DeY90,DeY92,Smi82,HLT96,Eva95}

\nocite{DDL87,DeV94,SDP89,HKM88,KHS87}

\nocite{RLC95,Eng91,DeY93,TLX96}

\newpage

\begin{deluxetable}{lccccccc}

\tablecaption{SF history parameters}

\tablehead{
\colhead{object} & \colhead{D}&
\colhead{$M_G$} & \colhead{$\nu$}& \colhead{$t_{inf}$} & 
\colhead{$M_{burst}$} & \colhead{$t_{burst}$} & \colhead{$t_e$} 
\\
\colhead{} & \colhead{Mpc} & \colhead{$[10^{10}M_\odot]$}  &
\colhead{[Gyr$^{-1}$]}& \colhead{[Gyr]} & \colhead{$[10^{10}M_\odot]$} &
\colhead{[Gyr]} & \colhead{[Gyr]}\\
\colhead{ (1)}& \colhead{(2)}&\colhead{(3)}&\colhead{(4)}&
\colhead{(5)}&\colhead{(6)}&\colhead{(7)}&\colhead{(8)}}
\startdata
M82    & 3.25 & $1.8$ &  $1.2$ & $9$   & $0.02$      & $12.95$     & $0.05$  \nl
NGC 6090  & 175 & $41$  &  $0.6$ & $9$   & $0.16$      & $12.95$     & $0.05$  \nl
ARP 220 & 115 & $23$  &  $0.3$ & $9$   & $2.5$       & $12.95$     & $0.05$  \nl
M51    & 9.6 & $15.5$&  $0.6$ & $4$   & \nodata     & \nodata     &\nodata  \nl
M100   & 20  & $20$  &  $0.75$& $4$   & \nodata     & \nodata     &\nodata  \nl
NGC 6946  & 6.7 & $12.5$&  $0.6$ & $5$   & \nodata     & \nodata     &\nodata  \nl
gE  & \nodata & $100$ &  $2.0$ & $0.1$ & \nodata     & \nodata     &\nodata  \nl
\enddata
\tablecomments{Parameters for the star formation history (Sec.
\ref{seccheevo}): (2) adopted distance; (3) final baryonic galaxy mass; (4)
SF efficiency; (5) infall timescale; (6) gas mass converted into stars during
the burst; (7) galaxy age at beginning of the burst (when included); (8)
e--folding time for SFR in the burst. All models have an age of 13 Gyr. In
the case of gE the SF has been stopped at 1.15 Gyr.}
\label{tabchi}

\end{deluxetable}

\begin{deluxetable}{lccccccc}
\small
\tablecaption{Geometric parameters}

\tablehead{
\colhead{object} & \colhead{$f_\mc$} & 
\colhead{$r_\mc$} &  
\colhead{$t_o$} & \colhead{$r_c^*$} & 
\colhead{$r_c^c$} & \colhead{$R_d$}& \colhead{$z_d$} \\
\colhead{}  & \colhead{} &
\colhead{[pc]}  & \colhead{[Myr]} &   \colhead{[kpc]} & \colhead{[kpc]} & 
\colhead{[kpc]} & \colhead{[kpc]} \\
\colhead{(1)}& \colhead{(2)}&\colhead{(3)}&\colhead{(4)}&\colhead{(5)}&
\colhead{(6)}&\colhead{(7)}&\colhead{(8)}
}
\startdata
M82    &   $0.08$     &$16$ & $57$   & $0.15$  & $0.2$  & \nodata  & \nodata \nl
NGC 6090  &   $0.005$ &$17$ & $18$   & $0.5$   & $1.0$  & \nodata  & \nodata \nl
ARP 220 &   $0.5$    &$10.6$& $50$   & $0.5$   & $0.5$  & \nodata  & \nodata \nl
M51    &   $0.7$      &$14$ & $8$    & \nodata & \nodata& $4.7$    & $0.4$   \nl
M100   &   $0.8$      &$15$ & $3$    & \nodata & \nodata& $5.0$    & $0.4$   \nl
NGC 6946  &   $0.6$   &$14$ & $2.5$  & \nodata & \nodata& $8.0$    & $1.0$   \nl
gE      &\nodata &\nodata&\nodata    & $0.4$   & $6.0$  & \nodata &\nodata\nl
\enddata

\tablecomments{
Parameters for the photometric model estimated from SED fitting: (2) fraction
of residual gas in MCs; (3) radius of MC; (4) parameter regulating
the escape of young stars from MCs (Eq.~\ref{equf}); (5)--(8) parameters for
the spatial distribution of stars, MCs, and cirrus (Eqs.~\ref{equdis} and
~\ref{equbul}). For disks the inclination angles have been taken from the
literature (see text) and are $i=20^o$ (M51), $i=30^o$ (M100), and $i=34^o$
(NGC 6946).}
\label{tabgeo}

\end{deluxetable}

\begin{deluxetable}{lcccccccc}
\small
\tablecaption{Derived quantities}

\tablehead{
\colhead{object}& \colhead{$<SFR>$} & \colhead{$M_{dust}$} & 
\colhead{$\tau_1^\mc$}  & \colhead{$\bar{\tau_1}$}& \colhead{$\bar{\tau_B}$} &
\colhead{$L_\mc$} & \colhead{$L_c$} &\colhead{$L_{Lyc}$}
\\
\colhead{ }     & \colhead{[$M_\odot/$yr]}& \colhead{$[10^7 M_\odot]$} &
\colhead{} & \colhead{ } & \colhead{ } &
\colhead{$10^{44}$ ergs/s} & \colhead{$10^{44}$ ergs/s} & \colhead{$10^{44}$ ergs/s}
\\
\colhead{(1)}& \colhead{(2)}&\colhead{(3)}&\colhead{(4)}&
\colhead{(5)}&\colhead{(6)}&\colhead{(7)}&
\colhead{(8)}&\colhead{(9)}
}
\startdata
M82    & $5.5$ &  $0.8$ & $25$ & $0.62$  & $1.30$  & 1.1 & 0.25& .09\nl
NGC 6090 & $68$  &  $45$& $24$ & $0.84$  & $1.37$  & 13  & 13  & 1.3\nl
ARP 220 & $580$ &  $30$ & $58$ & $1.67$  & $2.80$  & 100 & 5.7 & 8.7\nl
M51    & $6$   &$10.4$  & $33$ & $0.06$  & $0.28$  & 1.01& 0.90& 0.11\nl
M100   & $7$   &  $9.6$ & $29$ & $0.03$  & $0.15$  & 0.84& 0.85& 0.12\nl
NGC 6946 & $6$   &$10.4$& $33$ & $0.03$  & $0.12$  & 0.62& 0.53& 0.10\nl
gE     &\nodata & $0.15$&\nodata  & $0.00$  & $0.01$&\nodata& 0.02& 0.03\nl
\enddata
\tablecomments{A few quantities derived from the models: (2) SFR averaged
over the last $5\times 10^7$ yr, which is the time from burst onset for
starburst models; (3) total mass in dust in the galaxy;  (4) 1 $\mu$m
optical thickness of the MCs from the centre; (5) 1 $\mu$m `average' optical
thickness of the model, defined such as $(\textrm{observed flux}) =
(\textrm{dust--free flux} \times \exp (- \bar{\tau}))$ (6) same as (5) but
at 0.44 $\mu$m; (7) luminosities of the MCs, (8) of the diffuse dust and
(10) in the Lyman continuum {\it before dust absorption}.
}    
\label{tabqua}
\end{deluxetable}

\end{document}